\documentclass[twocolumn,showpacs,floats,floatfix,aps,pra]{revtex4}
\usepackage{amsfonts}
\usepackage{amssymb}
\usepackage{amsmath}
\usepackage{graphicx}
\usepackage{bm}
\usepackage{color}
\usepackage{epsfig}
\usepackage{ifthen}
\usepackage{color}
\usepackage{hyperref}

\begin{document}

\title{Efficient broadband sum and difference frequency generation with a
single chirped quasi-phase-matching crystal}
\author{A. A. Rangelov}
\affiliation{Department of Physics, Sofia University, James Bourchier 5 blvd., 1164
Sofia, Bulgaria}
\date{\today }

\begin{abstract}
We propose an efficient broadband frequency generation technique for two
collinear optical parametric processes $\omega_3=\omega_1+\omega_2$ and $%
\omega_4=\omega_1-\omega_2$. It exploits chirped
quasi-phase-matched gratings, which in the undepleted pump
approximation regime perform population transfer that is analogous
to adiabatic population transfer in a three-state ``vee'' quantum
system. The energy of the input fields is transferred
adiabatically either into $\omega_3$ or $\omega_4$ field,
depending on which of the two phase matchings occurs first by the
local modulation period in the crystal. One can switch the output
between $\omega_3$ and $\omega_4$ by inverting the direction of
the local modulation sweep, which corresponds to a rotation of the
crystal by angle $\pi$.
\end{abstract}

\maketitle


\section{Introduction}

Sum frequency generation (SFG) and difference frequency generation (DFG)
occur, when two input beams generate another beam with the sum (SFG) or the
difference (DFG) of the optical frequencies of the input beams \cite%
{Boyd,Saleh,Yariv}. In order to be efficient these processes
traditionally require phase matching \cite{Boyd,Saleh,Yariv},
which is usually difficult to achieve simultaneously for SFG and
DFG \cite{Boyd,Saleh,Yariv}.

Recently Suchowski \emph{et al.} \cite%
{Suchowski2008,Suchowski2009,Suchowski2011} used an aperiodically poled
quasi-phase-matching (QPM) crystal to achieve both high efficiency and large
bandwidth in SFG and DFG in the regime of undepleted pump approximation.
Their approach was based on ideas from rapid adiabatic passage in two-state
quantum systems \cite{Allen,Vitanov2001a,Shore}.

In this paper, this method is further extended to realize a potentially
highly efficient broadband SFG and DFG with a single crystal. This is
achieved by treating the two simultaneous collinear second-order parametric
processes $\omega _{3}=\omega _{1}+\omega _{2}$ and $\omega _{4}=\omega
_{1}-\omega _{2}$ in analogy to coherent population transfer in three-state
``vee'' quantum systems. The analogy to level crossings in atomic systems
\cite{Allen,Vitanov2001a,Shore} ensues from the linearly chirped QPM
gratings \cite{Lefort1,Lefort2,Arbore} that we use to this end.


\section{Background}

The SFG and DFG processes for a QPM crystal with susceptibility $\chi ^{(2)}
$ and local modulation period $\Lambda (z)$ are described by two sets of
nonlinear differential equations \cite{Boyd,Saleh,Yariv}
\begin{subequations}
\label{SFG}
\begin{align}
i\partial _{z}E_{1}& =\Omega _{1}E_{2}^{\ast }E_{3}e^{-i\Delta _{1}z}, \\
i\partial _{z}E_{2}& =\Omega _{2}E_{1}^{\ast }E_{3}e^{-i\Delta _{1}z}, \\
i\partial _{z}E_{3}& =\Omega _{3}E_{1}E_{2}e^{i\Delta _{1}z}.
\end{align}%
\end{subequations}
\begin{subequations}
\label{DFG}
\begin{align}
i\partial _{z}E_{1}& =\Omega _{1}E_{2}^{\ast }E_{4}e^{-i\Delta _{2}z}, \\
i\partial _{z}E_{2}& =\Omega _{2}E_{1}^{\ast }E_{4}e^{-i\Delta _{2}z}, \\
i\partial _{z}E_{4}& =\Omega _{4}E_{1}E_{2}e^{i\Delta _{2}z},
\end{align}
\end{subequations}
where $z$ is the position along the propagation axis, $c$ is the
speed of light in vacuum, and $E_{j}$, $\omega _{j}$ and $n_{j}$
are the electric field, the frequency and the refractive index of
the $j$-th laser beam, respectively. The coupling coefficients are
\begin{equation}
\Omega _{j}=\chi ^{\left( 2\right) }\omega _{j}/4cn_{j}\ \ \ (j=1,2,3,4),
\label{couplings}
\end{equation}%
while the phase mismatches for SFG and DFG processes are
\begin{subequations}
\begin{eqnarray}
\Delta _{1} &=&\omega _{1}n_{1}/c+\omega _{2}n_{2}/c-\omega _{3}n_{3}/c+2\pi
/\Lambda , \\
\Delta _{2} &=&n_{1}\omega _{1}/c+\omega _{2}n_{2}/c-\omega _{4}n_{4}/c+2\pi
/\Lambda .
\end{eqnarray}
\end{subequations}
We combine Eqs. \eqref{SFG} with Eqs. \eqref{DFG} to write a
system of differential equations that describes the two
simultaneously running processes
\begin{subequations}
\label{nonlinear system}
\begin{align}
i\partial _{z}E_{1}& =\Omega _{1}\left( E_{1}^{\ast }E_{2}e^{-i\Delta
_{1}z}+E_{1}^{\ast }E_{4}e^{-i\Delta _{2}z}\right) , \\
i\partial _{z}E_{2}& =\Omega _{2}\left( E_{1}^{\ast }E_{3}e^{-i\Delta
_{1}z}+E_{1}^{\ast }E_{4}e^{-i\Delta _{2}z}\right) , \\
i\partial _{z}E_{3}& =\Omega _{3}E_{1}E_{2}e^{i\Delta _{1}z}, \\
i\partial _{z}E_{4}& =\Omega _{4}E_{1}E_{2}e^{i\Delta _{2}z}.
\end{align}
\end{subequations}

\section{Adiabatic evolution of SFG and DFG in undepleted pump approximation}

\begin{figure}[t]
\centerline{\includegraphics[width=5cm]{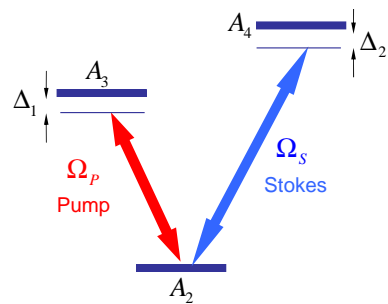}}
\caption{(Color online) The ``vee'' linkage pattern for the linearized
system (\protect\ref{three states system}), treated in analogy to the
``vee'' configuration of a three-state quantum system.}
\label{Fig1}
\end{figure}

\begin{figure}[t]
\centerline{\includegraphics[width=7.5cm]{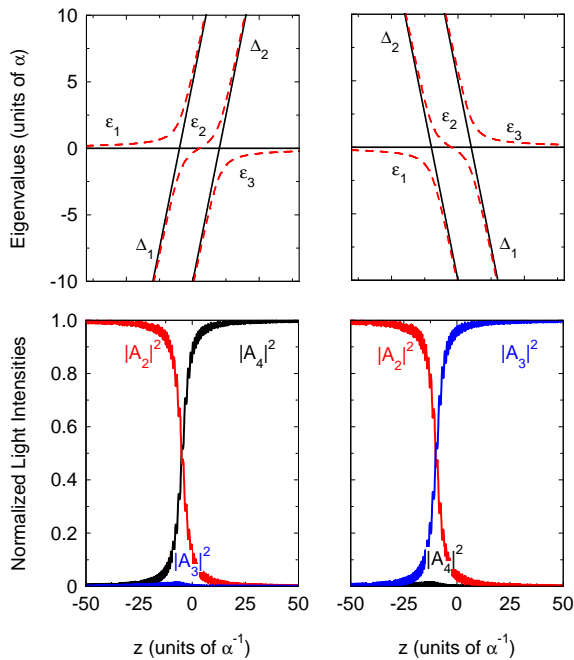}}
\caption{(Color online) Schematic evolution of frequency generation obtained
by numerical integration of Eqs.~(\protect\ref{three states system}) for $%
\protect\delta _{1}=-10\protect\alpha $, $\protect\delta _{2}=5\protect%
\alpha $ and $\Omega _{p}=\Omega _{s}=2\protect\alpha $. Left
frames: SFG with increasing phase mismatches. Right frames: DFG
with decreasing phase mismatches. Top frames: Diagonal elements
(solid lines) and eigenvalues (dashed lines) of the driving matrix
$\mathbf{H}$ of Eq. (\protect\ref{three states system}). Bottom
frames: normalized light intensities. The sign difference in Eq.
(\protect\ref{detunings}) between the left and the right frames
corresponds to z reversal (rotation of the crystal by angle
$\protect\pi $).} \label{Fig2}
\end{figure}

The coupled nonlinear equations (\ref{nonlinear system}) can be linearized
if we assume that the incident pump field $E_{1}$ is much stronger than the
other fields. In this case its amplitude is nearly constant (undepleted)
during evolution and as a result Eqs. \eqref{nonlinear system} are reduced
to a system of three linear equations,
\begin{subequations}
\label{three states system}
\begin{align}
i\partial _{z}\mathbf{A}(z)& =\mathbf{H}(z)\mathbf{A}(z),\quad  \\
\mathbf{H}& =\left[
\begin{array}{ccc}
\Delta _{1} & \Omega _{P} & 0 \\
\Omega _{P} & 0 & \Omega _{S} \\
0 & \Omega _{S} & \Delta _{2}%
\end{array}%
\right] ,
\end{align}
\end{subequations}
with
\begin{subequations}
\begin{align}
\Omega _{P}& =E_{1}\sqrt{\Omega _{2}\Omega _{3}}, \\
\Omega _{S}& =E_{1}\sqrt{\Omega _{2}\Omega _{4}}, \\
\mathbf{A}(z)& =[A_{3}(z),A_{2}(z),A_{4}(z)]^{T}, \\
A_{2}& =E_{2}/\sqrt{\Omega _{2}}, \\
A_{3}& =E_{3}e^{-i\Delta _{1}z}/\sqrt{\Omega _{3}}, \\
A_{4}& =E_{4}e^{-i\Delta _{2}z}/\sqrt{\Omega _{4}}.
\end{align}
\end{subequations}
Upon the substitution $z\rightarrow t$, Eq.~(\ref{three states
system}) becomes identical to the time-dependent Schr\"{o}dinger
equation for a
three-state quantum system in the rotating-wave approximation; the vector $%
\mathbf{A}(z)$ and the driving matrix $\mathbf{H}$ correspond to the quantum
state vector and the Hamiltonian, respectively. The diagonal terms of the
matrix $\mathbf{H}$, $\Delta _{1}$, $0$ and $\Delta _{2}$, correspond to the
detunings while the off-diagonal terms, $\Omega _{P}$ and $\Omega _{S}$,
correspond to pump and Stokes Rabi frequencies. We note that the quantity $|%
\mathbf{A}(z)|^{2}=|A_{3}(z)|^{2}+|A_{2}(z)|^{2}+|A_{4}(z)|^{2}$ is
conserved, analogously to the total population in a coherently driven
quantum system. If the energy is initially in the input electric field with
frequency $\omega _{2}$
\begin{equation}
\mathbf{A}=[0,A_{2},0],  \label{vector A}
\end{equation}%
then the three linear equations (\ref{three states system}) will form a
\textquotedblleft vee\textquotedblright\ pattern analogously to the
\textquotedblleft vee\textquotedblright\ configuration of a three-state
quantum system (see Fig. \ref{Fig1}). We assume that phase mismatches either
increase (sign $+$) or decrease (sign $-$) linearly along $z$%
\begin{subequations}
\label{detunings}
\begin{eqnarray}
\Delta _{1} &=&\delta _{1}\pm \alpha ^{2}z, \\
\Delta _{2} &=&\delta _{2}\pm \alpha ^{2}z,
\end{eqnarray}
\end{subequations}
which can be achieved, for example, by varying the local modulation period $%
\Lambda (z)$. For the sake of generality, we take hereafter $\alpha $ as the
unit of coupling and $1/\alpha $ as the unit of length. Therefore the three
eigenvalues of $\mathbf{H}$ cross at two different distances $z_{m}$ ($m=1,2$%
), thereby creating a crossing pattern in analogy with two parallel energies
crossed by a third, tilted energy in quantum physics \cite{Rangelov}. This
crossing pattern can be easily examined by the famous Landau-Zener-St\"{u}%
ckelberg-Majorana (LZSM) model \cite{Landau,Zener,Stuckelberg,Majorana},
which is the most popular tool for estimating the transition probability
between two states of crossing energies. This model assumes a constant
interaction of infinite duration and linearly evolving energies. Owing to
some mathematical subtleties, the LZSM model often provides more accurate
results than anticipated. Its popularity is further promoted by the extreme
simplicity of the transition probability expressions it derives. The LZSM
model has been extended to three and more levels by a number of authors. In
the Demkov-Osherov (DO) model \cite{Demkov,Kayanuma}, a single tilted energy
crosses a set of $N$ parallel energies. Our case, cf. Eqs. \eqref{three
states system} and Eqs. \eqref{detunings}, matches the DO model with two
parallel energies crossed by single tilted energy, as shown in the top
frames of Fig. \ref{Fig2}.

The proposed SFG and DFG are illustrated in Fig. \ref{Fig2}. The top frames
plot the eigenvalues of $\mathbf{H}$ ($\varepsilon _{1}$, $\varepsilon _{2}$
and $\varepsilon _{3}$) vs $z$. Initially only the $\omega _{2}$ field is
present (see Eq.(\ref{vector A})). If the evolution is adiabatic then there
are two possible paths that the system can follow (left and right frames).
If the phase match for the $\omega _{3}$ generation process occurs first
(left frames of Fig.~\ref{Fig2}), then the energy is passed into the $\omega
_{3}$ field. If instead first we observe the phase match for the $\omega _{4}
$ generation process (right frames of Fig.~\ref{Fig1}), then efficient
energy transfer to the $\omega _{4}$ field takes place.

The bottom frames of Fig. \ref{Fig2} show the evolution of the normalized
light intensities for the three possible frequencies $\omega _{2}$, $\omega
_{3}$ and $\omega _{4}$. The left and the right scenarios extend the
single-step adiabatic passage scenario for either SFG or DFG \cite%
{Suchowski2008,Suchowski2009,Suchowski2011}.

In the beginning and at the end each eigenfrequency $\varepsilon _{i}(z)$, $%
i=1,2,3$, coincides with one of the intrinsic diagonal terms of $\mathbf{H}$%
, while in between it is a superposition of these terms. In the adiabatic
limit, the system follows the eigenstate of $\mathbf{H}(z)$, which
asymptotically coincides with the initial state of the system.
Correspondingly, the frequency of the system at any instant of $z$ is the
frequency of this state, i.e. for the initial condition (\ref{vector A}) $%
\varepsilon _{1}(z)$. However, the composition of $\varepsilon _{1}(z)$ is
different for DFG and SFG because the order of the crossings differ.
Reordering the crossings can be easily done by $z$ reversal. Hence one can
achieve either SFG or DFG with a single chirped QPM crystal just by rotating
it on an angle $\pi $.

Next we turn our attention to the conditions needed for adiabatic evolution
in the two distinct cases of SFG and DFG. By applying the LZSM model \cite%
{Landau,Zener,Stuckelberg,Majorana}
\begin{equation}
p=1-\exp (-2\pi \Omega ^{2}/\alpha ^{2}),
\end{equation}%
we find that to obtain transition probability larger than $1-\epsilon $ we
must satisfy the following conditions at each crossing
\begin{equation}
\frac{\Omega }{\alpha }>\sqrt{\frac{\ln (1/\epsilon )}{2\pi }},
\label{adiabatic conditions}
\end{equation}%
where $\Omega =\Omega _{p}$ for SFG and $\Omega=\Omega _{S}$ for DFG. One
can readily verify that the conditions are satisfied for the parameters used
in Fig.~\ref{Fig2}.

As was shown earlier \cite{Suchowski2008,Suchowski2009,Suchowski2011} in
analogy with atomic physics, adiabatic implementation of both SFG and DFG
leads to robustness of the adiabatic approach against variations of the
parameters such as propagation distance, couplings and initial (final) phase
mismatches. Our approach shares the same robustness as previously proposed
adiabatic schemes, which includes stability to variation of crystal
temperature, wavelengths of the input electric fields, crystal length and
angle of incidence.

\section{Conclusion}

We have used the analogy between the time-dependent Schr\"{o}dinger equation
and the SFG/DFG equations in the undepleted pump approximation regime to
propose an efficient broadband SFG/DFG technique realized with a single
crystal. A local modulation period sweep along the light propagation creates
crossings in the phase matching between different parametric processes,
which in combination with adiabatic evolution conditions allow for both
efficient and robust SFG and DFG for the input frequencies. Chirped
adiabatic QPM gratings offer robustness against variations of the parameters
of both the crystal and the electric fields, which include the crystal
temperature, the wavelengths of the input electric fields, the crystal
length and the angle of incidence.

The present work can be viewed as a generalization of the idea of Suchowski
\emph{et al.} \cite{Suchowski2011}, however applied simultaneously for SFG
and DFG in a single crystal.

\acknowledgments
This work is supported by the European network FASTQUAST and the Bulgarian
NSF grants D002-90/08 and DMU02-19/09.



\begin{thebibliography}{99}
\bibitem{Boyd} R. W. Boyd, Nonlinear Optics, Academic, New York, 2007.

\bibitem{Saleh} B. E. A. Saleh and M. C. Teich, Fundamentals of Photonics ,
John Wiley \& Sons, New Jersey, 2007.

\bibitem{Yariv} A. Yariv and P. Yeh, Photonics: Optical Electronics in
Modern Communications, Oxford University Press, New York, 2007.

\bibitem{Suchowski2008} H. Suchowski, D. Oron, A. Arie, and Y. Silberberg,
Phys. Rev. A. 78 (2008) 063821.

\bibitem{Suchowski2009} H. Suchowski, V. Prabhudesai, D. Oron, A. Arie, and
Y. Silberberg, Opt. Express 17 (2009) 12731.

\bibitem{Suchowski2011} H. Suchowski, B. D. Bruner, A. Ganany-Padowicz, I.
Juwiler, A. Arie, and Y. Silberberg, Appl. Phys. B 105 (2011) 697.

\bibitem{Allen} L. Allen, J. H. Eberly, Optical Resonance and Two-Level
Atoms, Dover, New York, 1987.

\bibitem{Vitanov2001a} N. V. Vitanov, T. Halfmann, B. W. Shore, and K.
Bergmann, Annu. Rev. Phys. Chem. 52 (2001) 763.

\bibitem{Shore} B. W. Shore, Acta Phys. Slovaka 58 (2008) 243.

\bibitem{Lefort1} M. Charbonneau-Lefort, B. Afeyan, and M. M. Fejer, J. Opt.
Soc. Am. B 25 (2008) 463.

\bibitem{Lefort2} M. Charbonneau-Lefort, M. M. Fejer, and B. Afeyan, Opt.
Lett. 30 (2005) 634.

\bibitem{Arbore} M. A. Arbore, O. Marco, and M. M. Fejer, Opt. Lett. 22
(1997) 865.

\bibitem{Rangelov} A. A. Rangelov, J. Piilo, and N. V. Vitanov, Phys. Rev. A
72 (2005) 053404.

\bibitem{Landau} L. D. Landau, Physik Z. Sowjetunion 2 (1932) 46.

\bibitem{Zener} C. Zener, Proc. R. Soc. Lond. Ser. A 137 (1932) 696.

\bibitem{Stuckelberg} E. C. G. St\"{u}ckelberg, Helv. Phys. Acta 5 (1932)
369.

\bibitem{Majorana} E. Majorana, Nuovo Cimento 9 (1932) 43.

\bibitem{Demkov} Y. N. Demkov and V. I. Osherov, Zh. Eksp. Teor. Fiz. 53
(1967) 1589 [Sov. Phys. JETP 26 (1968) 916].

\bibitem{Kayanuma} Y. Kayanuma and S. Fukuchi, J. Phys. B 18 (1985) 4089.
\end{thebibliography}
\end{document}